\documentclass[aps,prl,showpacs,groupedaddress,twocolumn]{revtex4}
\usepackage{amsmath,amssymb}
\usepackage{graphicx}

\DeclareMathOperator{\sech}{sech}

\renewcommand{\Im}{\text{Im}}

\newcommand{\DLT}[1][ ]{\mathcal{D}_{L/T#1}}
\newcommand{\DT}[1][ ]{\mathcal{D}_{T#1}}
\newcommand{\DL}[1][ ]{\mathcal{D}_{L#1}}
\newcommand{\T}[1][ ]{\mathcal{T}_{#1}}
\newcommand{\jE}[1][ ]{j_{E#1}}
\newcommand{\jS}[1][ ]{j_{S#1}}
\newcommand{\jLT}[1][ ]{j_{L/T#1}}
\newcommand{\jL}[1][ ]{j_{L#1}}
\newcommand{\jT}[1][ ]{j_{T#1}}

\newcommand{\fL}[1][ ]{f_{L#1}}
\newcommand{\fT}[1][ ]{f_{T#1}}
\newcommand{\ET}[1][ ]{E_{T#1}}

\newcommand{\integral}[1][]{\def\ArgI{{#1}}\IntegralRelay}
\newcommand\IntegralRelay[3][]{ \int_{\ArgI}^{#1} #3 \mathrm{d}#2 }

\newcommand{\mum}{\mathrm{\mu{}m}}
\newcommand{\cmss}{\mathrm{\frac{cm^2}{s}}}
\newcommand{\muVK}{\mathrm{\frac{\mu{}V}{K}}}

\newcommand{\TT}[1][]{\left<\T[#1]\right>}
\newcommand{\MT}[1][]{M_{T#1}}
\newcommand{\ML}[1][]{M_{L#1}}
\newcommand{\MLT}[1][]{M_{L/T#1}}

\newcommand{\tanhe}[1][]{\fL[#1]^0}
\newcommand{\RT}{\tilde{R}}

\newcommand{\refdeltasharpening}{\ref{fig:delta-sharpening}a}
\newcommand{\refTcoupling}{\ref{fig:T0-coupling}b}
\newcommand{\refphase}{\ref{fig:phase}c}
\newcommand{\refjSandTT}{\ref{fig:jS-and-TT}d}

\newcommand{\Q}{Q}
\newcommand{\QNN}{\Q_{\mathrm{NN}}}
\newcommand{\QNS}[1][]{\Q_{\mathrm{NS#1}}}

\begin{document}

\title{Thermopower induced by a supercurrent in
  superconductor-normal-metal structures}

\author{Pauli Virtanen}
% \email[]{Pauli.Virtanen@hut.fi}
\author{Tero T. Heikkil\"a}
\email[]{Tero.T.Heikkila@hut.fi}
\affiliation{Low Temperature Laboratory, Helsinki University of
  Technology, P.O. Box 2200 FIN-02015 HUT, Finland.}

\begin{abstract}
We examine the thermopower $\Q$ of a mesoscopic normal-metal (N) wire
in contact to superconducting (S) segments and show that even with
electron-hole symmetry, $\Q$ may become finite due to the presence of
supercurrents. Moreover, we show how the dominant part of $\Q$ can be
directly related to the equilibrium supercurrents in the structure. In
general, a finite thermopower appears both between the N reservoirs
and the superconductors, and between the N reservoirs themselves. The
latter, however, strongly depends on the geometrical symmetry of the
structure.
\end{abstract}

\pacs{74.25.Fy, 73.23.-b, 74.45.+c}

\maketitle
%%%%%%%%%%%%%%%%%%%%%%%%%%%%%%%%%%%%%%%%%%%%%%%%%%%%%%%%%%%%%%%%%%%%%%%%%%%%%%%

%%%%%%%%%%%%%%%%%%%%%%%%%%%%%%%%%%%%%%% \section{Introduction}

%%% About the breaking of the Mott relation
Thermoelectric effects in electrical conductors typically result from
the asymmetry of the Fermi sea between the electron ($E>E_F$) and
hole-like ($E<E_F$) quasiparticles. This is illustrated by the Mott
relation~\cite{ashcroft} for the thermopower
\begin{equation}\label{eq:mott}
  \Q \equiv
  \left.
  \frac{\Delta V}{\Delta T}
  \right|_{I=0}
  =
  -
  \frac{\pi^2}{3} \, \frac{k_B^2 T}{e} \,
  \left.
  \frac{\mathrm{d}\ln{\sigma(E)}}{\mathrm{d}E}
  \right|_{E=E_F}
  \,.
\end{equation}
This relates the potential difference $\Delta V$ generated by the
temperature difference $\Delta T$ to the energy dependence of the
conductivity $\sigma$ due to the asymmetry above and below the Fermi
sea.

%%% What does this lead to.
In metals, the electron-hole asymmetry is governed by the
parameter $k_B T/E_F$ arising from the next-to-leading term in the
Sommerfeld expansion. At sub-Kelvin temperatures, this leads to a very
small $Q$, typically below 10~nV/K. However, recent
experiments~\cite{eom98,dikin02,parsons03} measuring the thermopower
in normal-metal wires connected to superconducting electrodes indicate
that it exceeds this prediction at least by an order of
magnitude, and, moreover, show that $Q$ oscillates
with the phase difference between the two superconducting contacts.

\begin{figure}
  \includegraphics{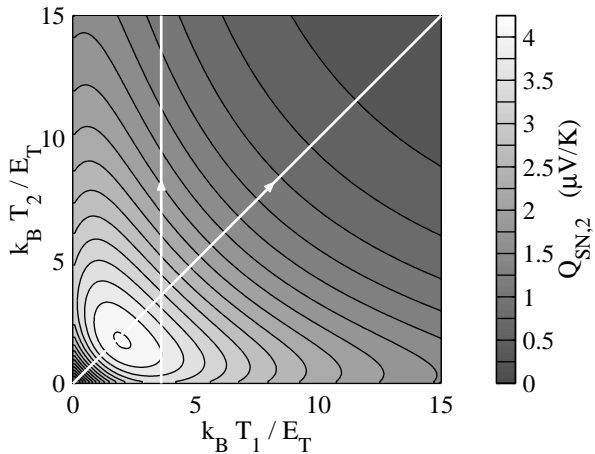}
  \caption{\label{fig:Qsn} N--S thermopower at different temperatures
    $T_1$ and $T_2$ of the normal electrodes, for a setup with $L_k$
    and $A_k$ assumed equal for all wires, and in the limit $\Delta
    \gg \ET,eV,k_B T$. The analytical
    approximation~\eqref{eq:scapprox} with
    $\T$-correction~\eqref{eq:Tcorrection} deviates at worst $15\%$
    from these results, but the difference is over 5\% only where
    $\QNS$ is small, i.e. $k_B T_1,k_B T_2\gtrsim 10~\ET$.  In all of
    the figures, $E_T=\hbar D/(L_3+L_4+L_5)^2$, and we plot the
    results for $\phi=\pi/2$ yielding a near-maximal supercurrent.}
\end{figure}

%%% What kind of an effect our effect is?
Mott relation is expected to fail in the presence of the
superconducting proximity effect when the geometrical symmetry in
the measured sample is broken~\cite{heikkila00}. Our aim is to
show that with nearby superconductors, normal-metal circuits can
show a thermoelectric effect independent of
electron--hole-symmetries, since the proximity effect couples the
temperatures to the potentials through the supercurrent. This
effect is at least two orders of magnitude larger than that
predicted by Eq.~\eqref{eq:mott} (c.f. Fig.~\ref{fig:Qsn}).

%%%%%%%%%%%%%%%%%%%%%%%%%%%%%%%%%%%%%%% \section{A qualitative result}

\begin{figure}
  \includegraphics{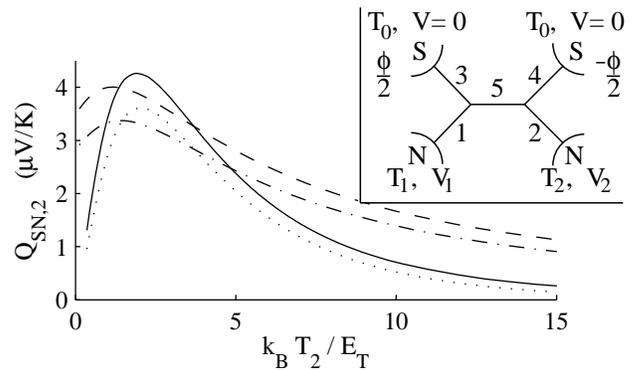}
  \caption{\label{fig:N-S}\label{fig:setup} N--S thermopower along the
    cross sections of Fig.~\ref{fig:Qsn}.  Solid: $\QNS$ at
    $T_1\approx T_2$. Dotted: Approximation~\eqref{eq:scapprox}. The
    correction~\eqref{eq:Tcorrection} accounts for most of the
    difference.  Dashed: $\QNS$ at $k_B T_1 = 3.6~\ET$ with varying
    $T_2$. Dash-dotted: the corresponding approximation. Inset: The
    setup under consideration: two superconducting (S) reservoirs with
    phase difference $\phi$ connected to two normal-metal (N)
    reservoirs through diffusive normal-metal wires. We assume that
    the lengths $L_k$, $k=1,\ldots,5$ of the wires satisfy $\xi_0
    \lesssim L_k \ll l_\phi,l_E$, where $\xi_0=\sqrt{\hbar
    D/(2\Delta)}$ is the superconducting coherence length, $l_\phi$
    the phase-coherence length and $l_E$ the energy-relaxation
    length. We also consider the wires as quasi-1D structures
    \cite{heikkila02scdos}.}
\end{figure}

%%% The qualitative picture: the N-S thermopower effect
We discuss the system shown in the inset of
Fig.~\ref{fig:setup}, with a supercurrent flowing between the two
superconducting elements. Our main result
(Eq.~\eqref{eq:scapprox}) states that the thermopower $\QNS$
between the N and S parts of the structure is
proportional to the difference in the supercurrents at the
temperatures $T_1$, $T_2$ of the two N electrodes. Moreover, we
obtain a similar result (Eq.~\eqref{eq:nnapprox}) for the
thermopower $\QNN$ between the normal parts in a geometrically
asymmetric structure. This can be understood phenomenologically as
follows. If $T_1 \ne T_2$, the
temperature-dependent~\cite{dubos01} equilibrium supercurrent
$I_S(T_1)$ in wire 3 is different from $I_S(T_2)$ in wire 4. (For
this qualitative picture, we approximate these wires to be at the
temperatures $T_1$, $T_2$.) Thus, a compensating effect must arise
to guarantee the conservation of currents. Should the N
reservoirs be kept at the same potential as the superconductors, a
quasiparticle current $I_{\text{qp}} \propto (I_S(T_1)-I_S(T_2))$
from them to the superconductors would balance the difference.
However, when no current is allowed to flow
in wires 1 and 2, a compensating N--S potential difference
$V_N-V_S \propto R (I_S(T_1) - I_S(T_2))$ is induced instead. The
induced potentials oscillate with phase differences, similarly to
the supercurrent, and may differ in the two N reservoirs,
especially in asymmetric structures.

%%%%%%%%%%%%%%%%%%%%%%%%%%%%%%%%%%%%%%% \section{Rigorous calculation}

%%% The rigorous calculation
In the following, we concentrate on the diffusive limit, and model
the setup with the Keldysh-Usadel equations~\cite{belzig99}. These equations assume electron-hole symmetry, under
which Eq.~\eqref{eq:mott} predicts a vanishing $Q$.

%%%%%%%%%%%%%%%%%%%%%%%%%%%%%%%%%%%%%%% \subsection{Usadel equations}

%%% The Usadel equations
The Keldysh-Usadel equations are formulated in terms of the
quasiclassical Green's functions (which are matrices in the
Keldysh-Nambu space), but here we use their
$\theta$-parametrization~\cite{belzig99}, for convenience.  It reduces
the problem to two sets of equations, the spectral equations and the
kinetic equations. In terms of the parameterizing functions $\chi$ and
$\theta$, the phase and the proximity effect strength, the spectral
equations in normal metals are
\begin{subequations}
  \label{eq:spectral}
  \begin{gather}
    D \nabla^2 \theta = -2 i E \sinh(\theta)
    + \frac{1}{2} D (\nabla \chi)^2 \sinh(2\theta) \,,
    \label{eq:spectral1}
    \\
    D \nabla\cdot \jE = 0,
    \quad \jE \equiv -\sinh^2(\theta) \nabla\chi \,.
    \label{eq:spectral2}
  \end{gather}
\end{subequations}
Here, the factor $D$ is the diffusion constant of the normal metal,
and $E$ is the energy with respect to the superconductor
potential. The kinetic equations are expressed using the symmetric and
antisymmetric parts, $\fT$ and $\fL$~\cite{belzig99}, of the electron
distribution function $f(E,\vec{r})$:
\begin{subequations}
  \label{eq:kin}
  \begin{align}
    D\nabla\cdot \jL &= 0, &
    \jL &\equiv \DL\nabla\fL - \T\nabla\fT + \jS\fT
    \label{eq:kinjL}
    \,,
    \\
    D\nabla\cdot \jT &= 0, &
    \jT &\equiv \DT\nabla\fT + \T\nabla\fL + \jS\fL
    \label{eq:kinjT}
    \,.
  \end{align}
\end{subequations}
The equations imply that the spectral current densities $\jL$ and
$\jT$ are conserved~(we neglect inelastic scattering). Thus the
observable charge and energy current densities,
\begin{align}
  \label{eq:currents}
  j_c = \frac{\sigma^N}{2 e} \integral{E}{\jT} \quad \text{and} \quad
  j_Q = \frac{\sigma^N}{2 e^2} \integral{E}{E \jL}\,,
\end{align}
are also conserved.

%%% The coefficients
The coefficients $\DL$, $\DT$, $\T$ and $\jS$ appearing in the kinetic
equations are obtained from the spectral equations:
\begin{subequations}\label{eq:coefficients}
  \begin{gather}
    \!\!\DLT\equiv\frac{1}{2}\left(1+|\cosh\theta|^2\mp|\sinh\theta|^2\cosh(2\Im[\chi])\right), \\
    \T \equiv\frac{1}{2}|\sinh\theta|^2\sinh(2\Im[\chi]),\quad
    \jS\equiv\Im[\jE].
  \end{gather}
\end{subequations}
Here, $\DL$ and $\DT$ are the local spectral energy and charge
conductivities, and $\jS$ is the spectral density of the
supercurrent~\cite{heikkila02scdos}. The factor
$\T$ arises in the formalism and has an effect on the thermopower.
The normal-state values of these coefficients are $\DL=\DT=1$ and
$\jS=\T=0$.

%%% Boundary conditions at nodes
At nodes of wires, assuming clean metallic contacts, the functions
$\theta$, $\chi$ and $f$ are continuous and Kirchoff-like ``spectral
current conservation laws''~\cite{nazarov99} imply that
$A\sigma^N\jLT$ and $A\sigma^N\jS$ are conserved. Here, $A$ is the
cross-sectional area of a wire and $\sigma^N$ the normal-state
conductivity.
%%% Boundary conditions at the reservoir interfaces
At clean metallic reservoir contacts most of the functions get their
bulk values~\cite{belzig99}.  However, for energies below the
superconducting energy gap $\Delta$, the valid boundary conditions at
superconductor interfaces are $\jL=0$, prohibiting the energy flow,
and $\fT=0$, assuming no charge imbalance in the superconductors.

%%%%%%%%%%%%%%%%%%%%%%%%%%%%%%%%%%%%%%% \subsection{Our analytical approach}

%%% General properties. What does couple?
The coefficients $\jS$ and $\T$ couple the energy and charge currents
together, and give rise to a finite thermopower. Moreover, these
coefficients oscillate with the phase difference in the system, and
thus the value of the thermopower should also oscillate. When there is
no phase difference, $\T=\jS=0$, and the thermopower vanishes.

%%% The Thouless energy
The energy scale of temperatures and potentials is specified by the
Thouless energy
\begin{equation}
  \ET \equiv \frac{\hbar D}{L^2}
  \approx \left\{
  \begin{array}{cc}
    13~\mu \text{V} & \; e \\
    0.15~\text{K}   & \; k_B
  \end{array}
  \right\} \;
  \frac{D/(200~\cmss)}{(L/\mum)^2}
  \,,
\end{equation}
corresponding to a wire of length
$L$~\cite{heikkila02scdos}. Moreover, $\ET$ of the link between the
superconductors is a natural energy scale for the spectral equations.
As long as $\ET\!\ll\!\Delta$, the results can be scaled to fit all
systems with similar ratios of wire lengths and areas.

%%% What thermopower means in our system? How we calculate it.
Since there are no general analytical solutions to the problem,
we solve the spectral equations numerically, and make a few
approximations to solve the kinetic equations. However, the data shown
in the figures is obtained numerically without any approximations.

%%% Derivation of the spectral current--distribution function
%%% relation, and necessary and additional approximations.
First, we note that the ``local potential'' $\fT$ is generally small
(as shown by the numerical results), as are the induced potentials at
the reservoirs. Thus, we can neglect the terms proportional to it in the
kinetic equation~\eqref{eq:kinjL}. Physically this means that we
mainly neglect the effect of supercurrent on the energy currents and
the temperatures. (If the potentials were large, the omitted term
would be the source for a Peltier-like effect~\cite{heikkila03}.) With
this approximation, we integrate the kinetic equations, which yields
the connection between the spectral current densities and the
distribution functions $\fL(x)$ and $\fT(x)$ at the ends of a wire of
length $L$:
\begin{subequations}
  \label{eq:approxjLT}
  \begin{align}
    \jL &= \frac{1}{L \ML} \, (\fL(L) - \fL(0)) \label{eq:approxjL}
    \,,
    \\
    \jT &= \label{eq:approxjT}
    \frac{1}{L \MT} \, (\fT(L) - \fT(0))
    + \jS \fL(0)
    \\
    &+ \frac{1}{L^2 \ML \MT} \!\left( \integral[0][L]{x}{\!\!\!\frac{\T}{\DL \DT}}
    + \jS \int_0^L \!\!\! \int_0^x \!\!\! \frac{\mathrm{d}x' \mathrm{d}x}{\DL(x') \DT(x)}
    \right)
    \notag
    \\
    &\quad\times (\fL(L) - \fL(0))
    \notag
    \,.
  \end{align}
\end{subequations}
Here $\MLT \equiv \frac{1}{L} \int_0^L \DLT^{-1}\,\mathrm{d}x$ are
the dimensionless spectral energy and charge resistances. To simplify
the final result, we also approximate $\DL=1$ in~\eqref{eq:approxjLT},
and $\DT=1$ in the latter term in~\eqref{eq:approxjT}, since the
variation in $\DL$ (away from superconductor interfaces) and $\DT$
with respect to the energy is smaller than that of the other
coefficients. Numerical results verify that this does not
affect the result crucially. The energy-dependent $1/\MT$ as a
coefficient for $\fT$ causes an important temperature dependence of
the conductance, so we retain it.

%%% How to proceed with derivation.
Using Eqs.~\eqref{eq:approxjLT} and the conservation of spectral
currents, we obtain a linear system of equations for the spectral
current densities. They can be solved with respect to the given
temperatures and potentials in the reservoirs, with different results
for $|E|<\Delta$ and $|E|>\Delta$, due to the different boundary
conditions. Next, we integrate over the energy to obtain the
observable current densities, after which we require the condition
$j_{c,1}=j_{c,2}=0$. To solve the resulting equations for the small
induced potentials $eV_1$ and $eV_2$, we linearize the distribution
functions with respect to them, and obtain a linear equation for the
potentials, which can then be solved.

%%%%%%%%%%%%%%%%%%%%%%%%%%%%%%%%%%%%%%% \section{Our results}

%%% Which is the most major contributor? The N-S thermopower.
If we proceed with the analytical approximation in the limit $\Delta
\gg \ET,eV,k_B T$ by neglecting $\T$ and the energy dependence of $\DT$,
we obtain the dominant term:
\begin{equation}\label{eq:scapprox}
  \begin{split}
    V_{1/2}^0
    =
    \frac{1}{2}
    \frac{R_5 (2 R_{4/3} + R_5) \;\; R_{3/4} ( I_S(T_1) - I_S(T_2) ) }%
         {(R_1 + R_2 + R_5)(R_3 + R_4 + R_5)}
    \,.
  \end{split}
\end{equation}
Here, $I_S(T)=(A \sigma^N/e) \integral[0][\infty]{E}{\jS\tanhe}$ is
the observable equilibrium supercurrent flowing in the system when all
parts are at the temperature $T$ and there are no potential
differences. Moreover, $\tanhe[,k] \equiv \tanh\left(E/(2 k_B
T_k)\right)$ is a linearized distribution function, and $R_k=L_k /
(A_k \sigma^N_k)$ are the normal-state resistances of the wires. Thus
a difference in equilibrium supercurrents due to the varying $f_L$
contributes significantly to the thermopower.

%%% TT corrections
Similarly, we can take the effect of $\T$ into account and obtain the
correction terms
\begin{align}\label{eq:Tcorrection}
  eV_{1/2}^1
  &=
  \frac{\mp R_{1/2}}{R_1+R_2+R_5} \,
  \integral[0][\infty]{E}{\!\!\!\left(\tanhe[,1]-\tanhe[,2]\right)\TT[1/2]\,}
  \\
  &\mp
  \frac{R_{3/4} R_5}{(R_1+R_2+R_5)R_{SNS}}
  \integral[0][\infty]{E}{\!\!\!\left(\tanhe[,1]-\tanhe[,2]\right) \TT[5]\, }
  \,,\notag
\end{align}
where $R_{SNS}=R_3+R_4+R_5$. Here, we denoted $\TT[k] \equiv
\frac{1}{L_k} \integral[0][L_k]{x}{\T[k]\,}$ (shown in
Fig.~\refjSandTT~compared with $\jS$).  The correction is necessary,
as it compensates for the fast decay of $V_{1/2}^0$ at high
temperatures $k_B T\gtrsim 10~\ET$, but it is not negligible even at
lower temperatures (see Fig.~\ref{fig:N-S}).

%%% Induced N-S themopower. Magnitude. Form.
The induced $\QNS[,2]\equiv V_2/(T_2 - T_1)\approx \QNS[,1]$ is
shown in Fig.~\ref{fig:Qsn}. The magnitude of $\QNS$ is of the order $\sim \muVK$ 
at highest, but there is also a strong temperature
dependence. Figure~\ref{fig:N-S} shows cross sections of
Fig.~\ref{fig:Qsn}, compared with the approximation~\eqref{eq:scapprox}.

\begin{figure}
  \includegraphics[width=\columnwidth]{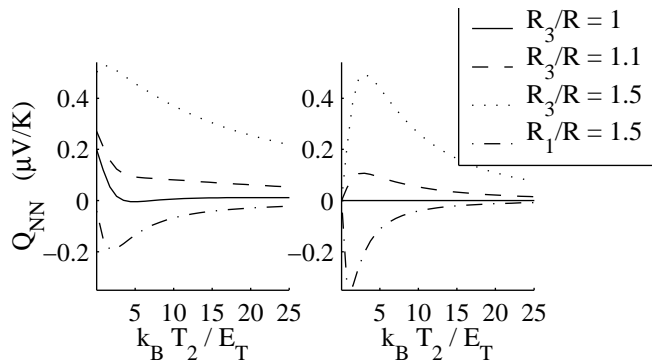}
  \caption{\label{fig:asymmetry-vs-DT} Left: N--N thermopower induced
    by setup asymmetry, at fixed $k_B T_1=3.6~\ET$. One wire (1 or 3)
    is assumed to have a different resistance than the others. Right:
    N--N thermopower with the same parameters, but with $T_1\approx T_2$. }
\end{figure}

%%% N--N asymmetry induced thermopower
We see that Eqs.~\eqref{eq:scapprox} and \eqref{eq:Tcorrection}
predict an induced N--N potential difference
\begin{equation}\label{eq:nnapprox}
  \Delta V \equiv V_2-V_1
  \approx (V_2^0 - V_1^0) + (V_2^1 - V_1^1)
    \,.
\end{equation}
In a left-right symmetric structure both terms vanish, but
$\QNN$ may still be finite since $\DT$ is energy-dependent:
\begin{align}\label{eq:dteffect}
  \Delta V
  &\approx
  \integral[0][\infty]{E}{
    \RT^{-1}
    \left(\frac{\sech^2\biglb(\frac{E}{2 k_B T_1}\bigrb)}{2 k_B T_1}
        - \frac{\sech^2\biglb(\frac{E}{2 k_B T_2}\bigrb)}{2 k_B T_2} \right)
  }
  \notag\\
  &\qquad\times (R_1 + (R_3 R_5)/(2 R_3 + R_5)) (V_1^0 + V_1^1)
  \,,
\end{align}
valid for a symmetric structure $R_1\!=\!R_2$, $R_3\!=\!R_4$. Here the
coefficient $\RT = \RT_1 + (\RT_3 \RT_5)/(2 \RT_3 + \RT_5)$, where
$\RT_k = \MT[,k] R_k$ are the spectral resistances of the wires. The
voltage~\eqref{eq:dteffect} can be understood to be caused by the
proximity-effect-induced temperature dependence of
conductances~\cite{charlat96}, which creates asymmetry in
resistances. However, in reality the asymmetry of the structure causes
likely a more significant effect (see Fig.~\ref{fig:asymmetry-vs-DT}),
especially at $T_1\approx T_2$, where Eq.~\eqref{eq:dteffect} predicts
$\Delta V \sim (T_2-T_1)^2$~\footnote{The $\DT$-effect may be
measurable by examining the symmetry of $\QNN$ around $T_1=T_2$.}.

%%% Phase oscillation
Equation~\eqref{eq:scapprox} implies that the thermopower should
oscillate as a function of the phase difference $\phi$ between
the two superconducting elements, because the equilibrium supercurrent
oscillates roughly as $\sin(\phi)$. Numerical simulations show (see
Fig.~\refphase) that also the exact solution oscillates similarly,
vanishing at $\phi=0$.

%%% On setup geometry dependence.
Besides changing the prefactors in
Eqs.~(\ref{eq:scapprox},\ref{eq:Tcorrection}), varying the resistances
of the wires from the near-symmetric presented in the figures changes
the behavior in the supercurrent \cite{heikkila02scdos} and the
coefficient $\T$. In general, large departures from such symmetry
decrease the potentials.

%%% Contributions above \Delta
A finite value for $\Delta$ causes two distinct modifications to the
thermopower. First, the coefficients~\eqref{eq:coefficients} are
modified, but changes are mostly only quantitative, e.g. sharpening of
peaks (see Fig.~\refdeltasharpening). Secondly, there is also a
contribution from energies $E>\Delta$, which couples the
superconductor temperature $T_0$ to the system. The latter effect is
weaker than those predicted by Eqs.~\eqref{eq:scapprox} and
\eqref{eq:Tcorrection} at least for $\Delta>30~\ET$. Although the
coupling of $T_0$ is weak, it induces finite potentials even for
$T_1=T_2$ (Fig.~\refTcoupling).

%%% Comparison with the experiment
Our predictions agree quantitatively with the experimental results
with the correct order of magnitude for both the linear
thermopower \cite{dikin02} and the temperature scale \cite{eom98}.
The flux dependence (antisymmetric about $\phi=0$ --- this holds
for the exact as well as the approximate solutions) is in accord
with most of the measurements in \cite{eom98,dikin02,parsons03}.
However, we cannot explain the symmetric oscillations with respect
to zero flux, seen in the "house" interferometers in
Ref.~\cite{eom98}. Moreover, the main result,
Eq.~\eqref{eq:scapprox} cannot describe a sign reversal of
$\QNS$~\cite{parsons03}, but there is no principal reason
forbidding such an effect in a suitable structure. Nevertheless,
further experiments are required to quantitatively demonstrate the
connection between the thermopower and the supercurrent.

\begin{figure}
  \includegraphics{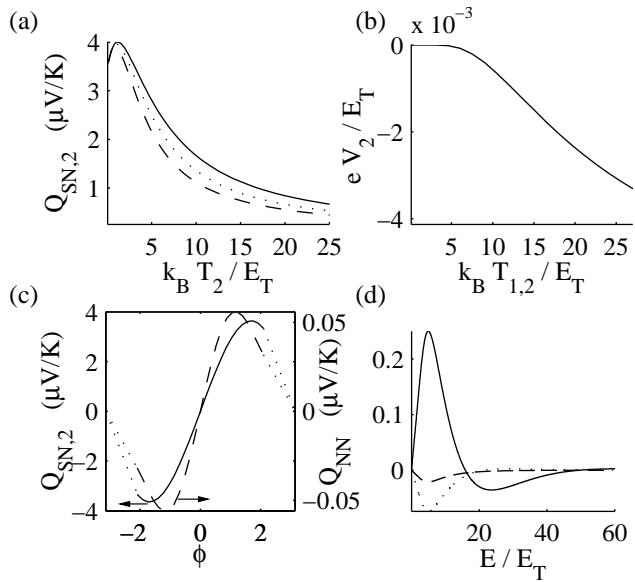}
  \caption{
    \label{fig:delta-sharpening}
    \label{fig:T0-coupling}
    \label{fig:phase}
    \label{fig:jS-and-TT}
    (a) N--S thermopower at $k_B T_1=3.6~\ET$ for various values of
    $\Delta$: $\infty$ (solid), $54~\ET$ (dotted), $27~\ET$
    (dashed). Superconductor temperature $T_0=3.6~\ET$ is fixed.  (b)
    Potential $e V_2$ induced due to the energies above $\Delta$, with
    $\Delta=27~\ET$. Here, $k_B T_0=3.6~\ET$ but $T_1=T_2$ vary.  (c)
    Phase oscillation of $\QNS$ (solid) and $\QNN$ (dashed), at $k_B
    T_1=4.5~\ET,\,k_B T_2=1.8~\ET$.  Due to a numerical convergence
    problem, there is no data for $|\phi|>2.1$.  (d) Spectral
    variables: $L_3 \jS$ (solid), $\TT[1]$ (dashed), $\TT[3]$
    (dotted).  In each figure, all the lengths $L_k$ and areas $A_k$
    are assumed equal.}
\end{figure}

%%% Reference to Volkov et al.
Results of similar type as presented in this Letter for the N-S
thermopower have been obtained for small temperature differences in
Refs.~\cite{seviour00,kogan02}, assuming high tunnel barriers at the
N-S contacts. However, the direct connection between the thermopower
and the supercurrents as in Eq.~\eqref{eq:scapprox} has not been
shown. Moreover, Ref.~\cite{kogan02} discusses a finite N--N
thermopower from the energies above $\Delta$. Our
results show that for an appreciable temperature difference between
the N reservoirs, this effect is washed out by the asymmetry effects,
at least for $\Delta\gtrsim30\ET$.

%%% Summary
In summary, we have obtained a relation linking the voltages
induced by a temperature difference to the supercurrent in a
mesoscopic structure. The phase-oscillating N--S thermopower is
mostly induced by the temperature dependence in the supercurrent,
and the N--N thermopower can be attributed to
left-right-asymmetries in a structure. These effects are
independent of electron-hole asymmetry, and can be much larger in
magnitude than the thermopower due to electron-hole symmetry
breaking.

%%%%%%%%%%%%%%%%%%%%%%%%%%%%%%%%%%%%%%%%%%%%%%%%%%%%%%%%%%%%%%%%%%%%%%%%%%%%%%%

\begin{acknowledgments}
We thank Jukka Pekola, Frank Hekking and Mikko Paalanen for helpful
discussions.
\end{acknowledgments}

\bibliography{main}

\end{document}